\documentstyle[aps,preprint]{revtex}
\tightenlines
\begin{document}
%\draft
\title{ On spectroscopy of  $\rho^\prime$, $\rho^{\prime\prime}$ and
$\omega^\prime$, $\omega^{\prime\prime}$ resonances}
\author{N.~N.~Achasov \footnote{Electronic address: achasov@math.nsc.ru}
and A.~A.~Kozhevnikov \footnote{Electronic address:
kozhev@math.nsc.ru}}
\address{Laboratory of Theoretical Physics, \\
Sobolev Institute for Mathematics \\ 630090, Novosibirsk-90,
Russia}
\date{Talk presented by A.~A.~Kozhevnikov at IX International Conference on Hadron
Spectroscopy {\it Hadron'01}, 25 August-1 September, 2001, IHEP,
Protvino, Russia} \maketitle \widetext
\begin{abstract}
Based on coupling constants extracted from fitting various data,
the selected branching ratios and full widths of $\rho^\prime$,
$\rho^{\prime\prime}$ and $\omega^\prime$, $\omega^{\prime\prime}$
resonances are calculated,  and some topics on the spectroscopy of
these states are  discussed.
\end{abstract}
\bigskip

The resonances $\rho^\prime\equiv\rho^\prime_1\equiv \rho(1450)$,
$\rho^{\prime\prime}\equiv\rho^\prime_2\equiv \rho(1700)$
$\left[I^G\left(J^{PC}\right)=1^+\left(1^{--}\right)\right]$ were
observed in the channels $\pi^+\pi^-$, $4\pi$, $\omega\pi$,
$\rho\eta$ of the reactions of $e^+e^-$ annihilation, $\tau$
lepton and $J/\psi\to\pi^+\pi^-\pi^0$ decays, photoproduction etc.
\cite{pdg}. The resonances
$\omega^\prime\equiv\omega^\prime_1\equiv\omega(1420)$,
$\omega^{\prime\prime}\equiv\omega^\prime_2\equiv\omega(1600)$
$\left[I^G\left(J^{PC}\right)=0^-\left(1^{--}\right)\right]$ were
observed in the channels $\pi^+\pi^-\pi^0$, $\omega\pi^+\pi^-$,
etc. The couplings with the states $K^\ast\bar K+$ c.c. and
$K^\ast\bar K\pi+$ c.c. are allowed. Possible indications come
from the reactions of $e^+e^-$ annihilation, photoproduction, etc.
\cite{pdg}. The resonances with the same $I^G$ are expected to be
strongly mixed via common decay modes. The partial width of each
specific mode is strongly energy dependent.

The approach to the treatment of the resonance mixing inspired by
field theory is used. It consists in the following steps,
assuming, for the sake of brevity, only two mixed states. First,
in case of zero width, zero mixing the propagator of the resonance
$R_a$, $a=1,2$, is  $1/d^{(0)}_a$. Second, in case of finite
width, zero mixing the propagator of the resonance $R_a$ becomes
$$\frac{1}{D^{(0)}_a}=\frac{1}{d^{(0)}_a}+\frac{1}{d^{(0)}_a}\Pi^{(0)}_{aa}\frac{1}{d^{(0)}_a}
+\cdots=\frac{1}{d^{(0)}_a-\Pi^{(0)}_{aa}},$$ Here
$\Pi^{(0)}_{aa}$  are the diagonal polarization operators. At
last, in case of  finite  width, nonzero mixing the propagator and
the nondiagonal 'polarization operator' $\Pi_{12}$ responsible for
the mixing are
$$\frac{1}{D_1}=\frac{1}{D^{(0)}_1}+\frac{1}{D^{(0)}_1}\Pi^{(0)}_{12}\frac{1}{D^{(0)}_2}
\Pi^{(0)}_{12}\frac{1}{D^{(0)}_1}+\cdots=
\frac{D^{(0)}_2}{D^{(0)}_1D^{(0)}_2-\Pi^{(0)2}_{12}}\equiv\left(G^{-1}\right)_{11},
$$ analogously for $1/D_2$, and $$
\frac{\Pi_{12}}{D^{(0)}_1D^{(0)}_2}=\frac{\Pi^{(0)}_{12}}{D^{(0)}_1D^{(0)}_2}
+\frac{[\Pi^{(0)}_{12}]^3}{[D^{(0)}_1D^{(0)}_2]^2} +\cdots
=\frac{\Pi^{(0)}_{12}}{D^{(0)}_1D^{(0)}_2-\Pi^{(0)2}_{12}}
\equiv\left(G^{-1}\right)_{12} ,$$ where $$ G(s)=\left(
\begin{array}{cc}D^{(0)}_1&-\Pi^{(0)}_{12}\\
-\Pi^{(0)}_{12}&D^{(0)}_2
\end{array}\right)\;$$ is the matrix of inverse propagators.
The physical states and their couplings   are obtained by the
diagonalization of $G^{-1}$. Hereafter $D^{(0)}_a\equiv
D^{(0)}_a(s)=m^{(0)2}_a-s-i\sqrt{s}\Gamma^{(0)}_a(s),$
$\Pi^{(0)}_{ab}\equiv\Pi^{(0)}_{ab}(s)=\mbox{Re}\Pi^{(0)}_{ab}+i\mbox{Im}\Pi^{(0)}_{ab}(s).$
By unitarity,
$\mbox{Im}\Pi^{(0)}_{aa}(s)=\sqrt{s}\sum_i\Gamma^{(0)}_{a\to i}
(s)$, $\mbox{Im}\Pi^{(0)}_{ab}(s)=\sqrt{s}\sum_i\Gamma^{(0)}_{a\to
i}(s) \frac{g^{(0)}_{R_bi}}{g^{(0)}_{R_ai}}$, where
$\Gamma^{(0)}_{a\to i}$ is the partial width of the decay $R_a\to
i$. Index (0) refers to the unmixed state. The masses $m^{(0)}_a$,
coupling constants $g^{(0)}_{R_ai}$, and $\mbox{Re}\Pi^{(0)}_{ab}$
assumed to be constants should be determined from experiment. The
generalization to the case of arbitrary number of states is
straightforward. The amplitude for the process $i\to R_1+R_2\to f$
is now
$$a_{fi}=(g^{(0)}_{fR_1},g^{(0)}_{fR_2})G(s)^{-1}\;\left(\begin{array}{c}g^{(0)}_{R_1i}
\\g^{(0)}_{R_2i}\end{array}\right)\;.$$
There are two essential points related with the resonance  mixing
and the dependence of partial widths on energy. First, the mass
shift of resonances due to their mixing is obtained by  the
condition of vanishing of the real part of det$G$. More important
is  the shift of the resonance position due to energy dependent
width. In the case of  the single resonance $R$ with the bare mass
$m_R$ produced in the reaction $e^+e^-\to R\to f$ whose cross
section can be written as $$ \sigma(s)=12\pi
m^3_R\Gamma_{Rl^+l^-}(m_R)g^2_{Rf} {s^{-3/2}W_{Rf}(s)\over
(s-m^2_R)^2+s\Gamma^2_R(s)}, $$ where $W_{Rf}$ is the phase space
volume of final state, the peak position is found from the
vanishing of derivative of $\sigma(s)$ with respect to $s$. At
sufficiently slow varying phase space the peak shifts to $$
s_R\approx
m^2_R-{1\over2}m^2_R\Gamma^2_R\frac{d}{ds_R}\left[\ln\left(s_R^{-3/2}W_{Rf}(s_R)\right)\right]
<m^2_R.$$

The data  are fitted  in the framework of the three-resonance
scheme \cite{donach} for each channel where the indication on the
specific resonance exists, see
\cite{snd99,snd01,ach97,ach98,ach00} and references therein
\footnote{We have used the earlier SND data \cite{snd99} on the
reaction $e^+e^-\to\pi^+\pi^-\pi^0$. Now new SND data \cite{snd01}
on this reaction are available. The question of whether these new
data can be described in the three resonance scheme with the only
$\omega^\prime_{1,2}$ states in addition to $\omega(782)$ or  new
states should be invoked is under study.}. The $V^\prime_{1,2}VP$
coupling constants extracted from the fits turned  out to be in
the intervals $\vert g_{\rho^\prime_1\omega\pi}\vert=10-18$
GeV$^{-1}$, and $\vert g_{\rho^\prime_2\omega\pi}\vert=2-13$
GeV$^{-1}$. Qualitatively, the relation $\vert
g_{\rho^\prime_{1,2}\omega\pi}\vert\sim\vert
g_{\omega^\prime_{1,2}\rho\pi}\vert$ was found to be satisfied.
$V^\prime_{1,2}VP$ coupling constants are not  suppressed as
compared to the $VVP$ ones. Coupling constants with multi-particle
states such as $g_{\rho^\prime_{1,2}\rho\pi^+\pi^-}$,
$g_{\omega^\prime_{1,2}\omega\pi^+\pi^-}$, as well as analogous
coupling containing strange mesons in final states, should also be
included. Taking $g_{\rho^\prime_{1,2}\omega\pi}\sim
g_{\omega^\prime_{1,2}\rho\pi} \simeq10$ GeV $^{-1}$, and
$m_{\rho^\prime_1}\approx m_{\omega^\prime_1}=1400$ MeV, one can
estimate the partial widths to be
$\Gamma_{\rho^\prime_1\to\omega\pi}\sim280\mbox{ MeV}$,
$\Gamma_{\omega^\prime_1\to\rho\pi}\sim820\mbox{ MeV}$.
 Analogously, assuming that
$m_{\rho^\prime_2}\approx m_{\omega^\prime_2}=1750$ MeV, one finds
$\Gamma_{\rho^\prime_2\to\omega\pi}\sim880\mbox{ MeV}$,
$\Gamma_{\omega^\prime_2\to\rho\pi}\sim2600\mbox{ MeV}$. $VP$
decay modes are not the only ones to which heavy resonances can
decay, and partial widths of the decay into many particle final
states are found to be of the same order or even larger as
compared to $VP$ ones. The results of the calculation of selected
branching ratios and total widths are presented in Tables
\ref{tab1}, \ref{tab2} and \ref{tab3}.
 One can see that the resonances
$\rho^\prime_{1,2}$ and $\omega^\prime_{1,2}$ are broad.

Our conclusions are as follows.
\begin{itemize}
\item One should be careful in attributing the specific peak or
structure in the cross section to the specific spectroscopy state,
because  the large width, the rapid growth of the phase space with
the energy increase, and the mixing among the resonances  result
in the shift of the visible peaks in the cross sections. For
example, large width of the $\omega^\prime_2$ resonance results in
the shift of its peak in the energy dependence of the
$e^+e^-\to\omega\pi^+\pi^-$ reaction cross section to $\simeq1680$
MeV from the bare mass $\simeq2000$ MeV \cite{ach98}.
\item $\rho(1300)$ state observed by LASS  team \cite{lass} in the reaction
$K^-p\to\pi^+\pi^-\Lambda$, revived an old discussion concerning
the possible existence of the $\rho(1250)$ meson, in addition to
the $\rho(1450)$ claimed to be observed in $e^+e^-$ annihilation.
The results presented in Table \ref{tab1}  show that the
corresponding peak should be attributed to the same state
$\rho(1450)$ as that presented in Reviews of Particle Physics
\cite{pdg}. \item The state $\omega(1200)$ observed  by  SND team
\cite{snd99,snd01} in the reaction $e^+e^-\to\pi^+\pi^-\pi^0$  is,
in our opinion, the same state as
$\omega(1420)\equiv\omega^\prime_1$ presented in PDG \cite{pdg}.
\item  The states $\rho^\prime_{1,2}$ and $\omega^\prime_{1,2}$
turn out to be the broad  resonance structures, as if the
conventional quark picture of them as the radial excitations is
implied. The present results, having in mind their significant
uncertainties, do not contradict to the assignment of
$\rho^\prime_1$ and $\omega^\prime_1$ resonances to the state
$2^3S_1$.  In the meantime, the central values of the
$\rho^\prime_2$ and $\omega^\prime_2$ widths are large, which
contradicts to the assigning them to the state $1^3D_1$ predicted
by Gogfrey and Isgur \cite{godfrey} to be relatively narrow.
However, large errors prevent one from drawing final conclusion.
\item The
very large widths of the resonances  may indirectly evidence in
favor of some nonresonant contributions to the amplitudes. The
accuracy of the existing data is still poor either to isolate such
contributions reliably or to specify their form precisely.
Considerable increase of experimental statistics hopefully
accessible at  VEPP-2000 collider aimed at the study of the energy
range of $e^+e^-$ annihilation from 1 to 2 GeV will help in
resolving the above issues.
\end{itemize}

\begin{table}
\caption{Masses, total widths (in the units of MeV), leptonic
widths (in the units of keV), and selected branching ratios
(percent) of  $\rho^\prime_1$ resonance,  calculated using the
coupling constants extracted from the fits \protect\cite{ach97} of
the specific reaction. The symbol $\sim$ means that only central
value is given, while  error exceeds it considerably. The
$\rho^\prime_1$ resonance does not reveal itself in the
$e^+e^-\to\omega\pi^0$ reaction and $\tau^-$ decay. Other
branching ratios can be found in Ref.~\protect\cite{ach00}.}
\begin{tabular}{lllll}
reaction &$e^+e^-\to\pi^+\pi^-$ &$e^+e^-\to2\pi^+2\pi^-$&
$e^+e^-\to\pi^+\pi^-2\pi^0$&$K^-p\to\pi^+\pi^-\Lambda$\\
$m_{\rho^\prime_1}$&$1370^{+90}_{-70}$&$1350\pm50$&
$1400^{+220}_{-140}$&$1360^{+180}_{-160}$\\
$B_{\rho^\prime_1\to\pi^+\pi^-}$&$1.1\pm1.1$&$\sim1.4$&
$\sim8.0$&$\sim0.7$\\
$B_{\rho^\prime_1\to\omega\pi^0}$&$86.5\pm41.5$&$93.6\pm60.0$&
$77.8\pm62.2$&$93.3\pm82.7$\\
$B_{\rho^\prime_1\to4\pi}$&$\sim6.8$&$\sim0.2$&$\sim7.2$&
$\sim0.7$\\ $\Gamma_{\rho^\prime_1\to
l^+l^-}$&$6.4^{+1.2}_{-1.4}$&$5.4^{+2.6}_{-1.8}$&$6.3^{+3.3}_{-2.5}$
&$-$\\
$\Gamma_{\rho^\prime_1}$&$763\pm500$&$\sim518$&$\sim970$&$\sim460$\\
\end{tabular}
\label{tab1}
\end{table}
\begin{table}
\caption{The same as in Table \protect\ref{tab1}, but in the case
of the $\rho^\prime_2$ resonance. The latter  does not reveal
itself in the $K^-p\to\pi^+\pi^-\Lambda$ reaction. }
\begin{tabular}{llllll}
reaction&$e^+e^-\to\pi^+\pi^-$&$e^+e^-\to\omega\pi^0$
&$e^+e^-\to2\pi^+2\pi^-$&
$e^+e^-\to\pi^+\pi^-2\pi^0$&$\tau^-\to(4\pi)^-\nu_\tau$\\
$m_{\rho^\prime_2}$&$1900^{+170}_{-130}$&$1710\pm90$&$1851^{+270}_{-240}$&
$1790^{+110}_{-70}$&$1860^{+260}_{-160}$\\
$B_{\rho^\prime_2\to\pi^+\pi^-}$&$\sim0$&$\sim0$&$\sim1.2$&$\sim0.4$&
$1.5\pm1.4$\\
$B_{\rho^\prime_2\to\omega\pi^0}$&$\sim16.7$&$22.3\pm8.0$&$13.4\pm3.9$&$31.0\pm18.6
$&$18.9\pm2.8$\\$B_{\rho^\prime_2\to4\pi}$&$\sim68.5$&$61.2\pm7.8$&$74.0\pm32.1
$&$45.0\pm18.0$&$62.6\pm5.0$\\$\Gamma_{\rho^\prime_2\to
l^+l^-}$&$1.8\pm1.5$&$5.2\pm1.5$&$4.02^{+0.28}_{-0.27}$&$4.5\pm1.3$&
$9.3\pm0.6$
\\$\Gamma_{\rho^\prime_2}$&$\sim303.9$&$1886\pm613$&$3123\pm296$&$3151\pm1281$&
$3255\pm388$\\
\end{tabular}
\label{tab2}
\end{table}
\begin{table}
\caption{Masses, total widths (in the units of MeV), leptonic
widths (in the units of eV), and branching ratios (percent) of the
$\omega^\prime_{1,2}$ resonances,  calculated using the coupling
constants extracted from the fits \protect\cite{ach98} of the
specific channel of $e^+e^-$ annihilation.  The symbol $\sim$
means that only central value is given, while  error exceeds it
considerably.} \label{tab3}
\begin{tabular}{llllll}
channel&$\pi^+\pi^-\pi^0$&$\omega\pi^+\pi^-$&$ K^+K^-$&$
K^0_SK^\pm\pi^\mp$&$K^{\ast 0}K^\mp\pi^\pm$\\
$m_{\omega^\prime_1}$&$1430^{+110}_{-70}$&$\sim1400$&$\sim1460$&$\sim1500$&$\sim1380$\\
$B_{\omega^\prime_1\to3\pi}$&$\sim21.6$&$\sim8$&$\sim67$&$\sim96$&$\sim34$\\
$B_{\omega^\prime_1\to K^\ast\bar
K+cc}$&$\sim0.2$&$\sim0$&$\sim1$&$\sim4$&0\\
$B_{\omega^\prime_1\to K^\ast\bar
K\pi}$&$\sim0$&0&0&0&0\\$B_{\omega^\prime_1\to\omega\pi^+\pi^-}$&$\sim78.2$&$\sim92$&
$\sim31.2$&$\sim0$&$\sim65.8$\\$\Gamma_{\omega^\prime_1\to
l^+l^-}$&$144^{+94}_{-58}$&$\sim0.2$&$\sim8$&$\sim8$&$\sim48$\\
$\Gamma_{\omega^\prime_1}$&$\sim903$&$\sim129$&$\sim173$&$\sim1252$&$\sim112$\\
$m_{\omega^\prime_2}$&$1940^{+170}_{-130}$&$2000\pm180$&$1780^{+170}_{-300}$&$\sim2120$&
$1880^{+600}_{-1000}$\\$B_{\omega^\prime_2\to3\pi}$&$\sim22.1$&$\sim34.2$&$\sim88.8$&
$\sim91.2$&$\sim60.1$\\$B_{\omega^\prime_2\to K^\ast\bar
K+cc}$&$\sim3.5$&$\sim5.8$&$\sim11.2$&$\sim15.8$&$\sim8.9$\\$B_{\omega^\prime_2\to
K^\ast\bar
K\pi}$&$\sim68.2$&$\sim53.4$&0&0&$\sim30.9$\\$B_{\omega^\prime_2\to\omega\pi^+\pi^-}$&
$\sim6.2$&$\sim6.6$&0&0&$\sim0$\\
$\Gamma_{\omega^\prime_2l^+l^-}$&$109^{+58}_{-46}$&$531\pm225$&0&$\sim189$&$1162\pm922$\\
$\Gamma_{\omega^\prime_2}$
&$\sim14000$&$\sim5757$&$\sim2420$&$\sim9854$&$\sim13820$\\
\end{tabular}
\end{table}--
\end{document}